# Impacting the bioscience progress by backporting software for Bio-Linux

Sasa Paporovic

sasa.paporovic@fh-bielefeld.de

v0.9

## What is Bio-Linux and what is it good for - also its drawbacks:

If someone says to use or to have a Linux this is correct as like it is imprecise. It does not exist a Linux as full functional operating system by itself. What was originally meant by the term Linux was the operating system core[1]. The so called kernel, or in a case of a Linux operating system the Linux kernel. It is originally designed and programmed by Linus Torvalds, who is also today the developer in chef or to say it with his words, he is the "alpha-male" of all developers[2].

Anyway, what we have today are Distributions[3]. It has become common to call them simply "a Linux". This means that there are organizations out there, mostly private, some funded and some other commercial, which gather all what is needed to design around the Linux kernel a full functional operating system. This targets mostly Software, but also web and service infrastructure. Some of them have a history that is nearly as long as the Linux kernel is alive, like Debian. Some others are younger like Ubuntu and some more others are very young, like Bio-Linux[4].

The last Linux, the Bio-Linux, especially its latest version Bio-Linux 7 we are focusing here[5].

In year 2006 Bio-Linux with the work of Tim Booth[42] and team gives its rising[6] and provide an operating system that was and still specialized in providing a bioinformatic specific software environment for the working needs in this corner of bioscience.

## A software environment?

Yes, as seen above a Linux is more than only its kernel(core of the system).
A Linux of any distribution incorporates, and in this point it is very different from a Windows, all available Software for this distribution.

This is done by a preparation step of the developers of the distribution that is called packaging[7]. In this package process the source code of a software will be gathered and compiled specific for all components that should be integrate in the finished Linux Distribution.
The packaging process will produce a package of the software that should be and is mostly capable of to be installed smoothly in the Linux.

This is done for every software that should be available for the finalized Distribution/Linux version, so that with the time huge software pools called repositories have been designed, some of them with several 10.000 programs in it[8].

Every Distribution installation is capable to get the available software from central servers on which the software pool reside and install it from there via package management tools like dpkg or rpm[9]. Installation of software without using the prepared packages from the central servers is still possible, but except for very simple programs this may require remarkable skills especially to keep sure to render the Linux not over time in an unusable state.

So, packaging is for trustful obtaining of software (every package is cryptographic signed) and easy installation.

But, this forces Linux developers to make a much more intensive job, then the developers at Microsoft have. Microsoft employers only have to orchestrate some hundred programs for a good running operating system they sell. Every additional software is primary an issue for the user/customer and with it the most problems that an external software may causing to the heart of a Windows.(e.g incompatible .dll files)[10].

Linux developers have taken over by their own choice the much harder job.
With the process of packaging, which is an extended form of system integration, they orchestrate the functionality for and with several 10.000 packages, respective software[11].

In principle it is thinkable to do this continuously without deadlines and heartbreaks of the package/software integration, so that all the time the latest software and software versions are available.
This is called a rolling release[12]. There are some approaches to do this, and one of the more famous ones is "Arch Linux"[13], which do time line independent integration of the latest software.

But, you know, nothing you can get without a price. Bringing in the latest software into a productive installation can bring the system in unstable states and can cause subtle and only hard foreseeable errors within the system.
As consequence the most Linux distributions have decided not to perform a rolling release for smoothness integration of software.

Instead they have development phases in so called release cycles with a defined end and Linux versions as result. For example Ubuntu[14], which will also be handled here.

In every development cycle new software is packaged and integrated, which means that one of the development objectives is to stabilize the imported software.
Logically in some point during the development cycle there is a import stop of new software or new software versions. After this point no new software will be incorporated for this specific release version.

After the import freeze the stabilization of the whole software canon for this release version begins and you see as consequence that every release version has a fixed software canon, that without a little number of exceptions will not be changed anymore until finalizing of development and also afterwards when it is final/stable.

So, dependent on the development time, between the availability of one final version and the next final version there might be huge time gabs. In this time gabs no newer software beside the freezed software canon will be available for the installations made by the users of the latest final version[15].

**How this all affects Bio-Linux?**

Simply Bio-Linux is also affected by a release cycle and with this the final releases of Bio-Linux will not have the latest bioinformatic software on board.

*What is and could be done to get around this problem?*

To obtain an answer let us see where the source of all bioinformatic software in Bio-Linux 7 is:

Mostly it begins on Debian Linux. There is a team that call itself Debian-Med[16]. Andreas Tille[41] is one of the most active members. They decide in the best of there knowledge what bioinformatic software is good for Debian and, properly more remarkable, they do the packaging and system integration job[17,18].

*And how comes Ubuntu in the game?*

Ubuntu is a bit parasitic to Debian. Oh, let me correct that. It is a symbiont.
They derive the Software for the actual development branch of Ubuntu from the actual testing or unstable branch of Debian[19]. The other way around Ubuntu provide problem information and bug fixes for the Debian project. Additional many Ubuntu developers are also Debian developers(or the other way around)[14].
To user terms of genealogy: Debian is the mother and Ubuntu the daughter.

*And what is the relation to Bio-Linux?*

Tim Boot and team do the same as the Ubuntu people with Debian. They derive, but not from Debian direct. Instead they derive from Ubuntu long term supported Versions, so called Ubuntu LTS. The latest derivings were Ubuntu10.04 for Bio-Linux 6 and Ubuntu12.04 for Bio-Linux 7. With this Debian is the grandmother, Ubuntu the mother and Bio-Linux the daughter.

*So, to summaries the game:*

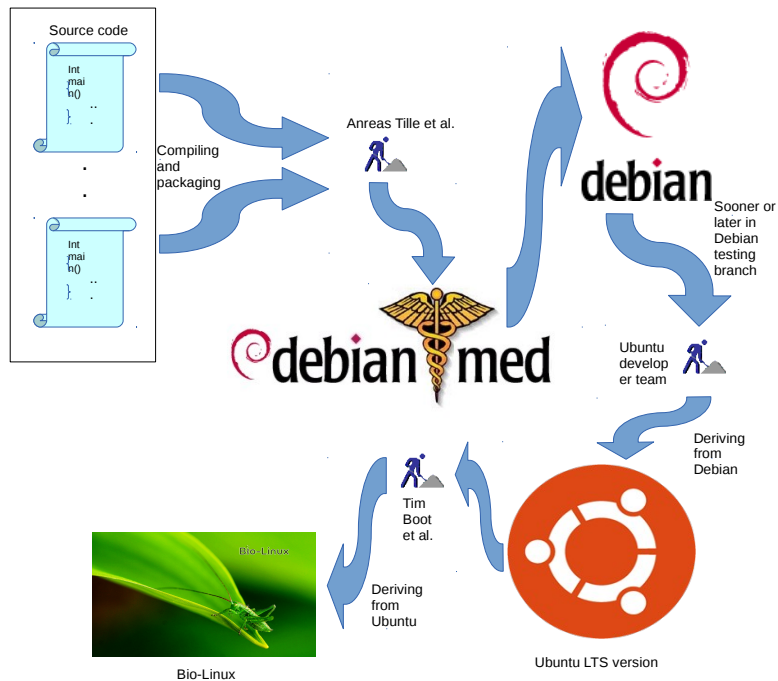

The source code of any bioinformatic software for Debian-Med is selected, compiled and packaged by Andreas Tille[41] and the Debian-Med Team. Sooner or later the software hits the testing branch of Debian-Linux and is derived from there into Ubuntu LTS(Long term supported version) of Ubuntu-Linux. Tim Booth[42] and the Bio-Linux team derive again the actual Bio-Linux version from the actual Ubuntu LTS version.

**The drawback of the workflow:**

There are only every two(even) years Ubuntu LTS-Versions(Ubuntu6.06 in 2006, Ubuntu8.04 in April 2008, Ubuntu10.04 in April 2010, Ubuntu12.04 in April 2012 and the next will be 14.04 in April 2014[20].

*Figure1: The software flow between different Linux distributions is shown and with this their genealogy. Also the involved workgroups are named. Icons and images are from differen online sources[16,37,38,39] The graphic is designed with LibreOffice 4.1.2 on Ubuntu13.10.*

It is not good to get only every two years fresh scientific software/software versions, which is the consequence of the fact that Bio-Linux is derived only on the LTS versions and their 2 year release cycle. The latest scientific developments are not available in this way, which slows down the scientific progress.
To circumvent this gab, the Bio-Linux team is incorporating additional software in Bio-Linux and bring new software version on over their own repository. It is enabled by default in Bio-Linux.
They give also back to Ubuntu and Debian and prepare with this also the next versions[21].

*Is this now all?*

No, there is still a other bridge on which new software could migrate into Bio-Linux.
Let us have a look.

**The backporting bridge for getting over the 2 year gab.**

*What is a backport?*

Your Bio-Linux7-System gather its software from different servers/sources, called repositories[8]. Some of them are the original repositories for Ubuntu12.04, from which Bio-Linux is derived. One other repository already mentioned above is direct from the Bio-Linux team and provide the software that makes Bio-Linux to Bio-Linux. It could be viewed at:

http://nebc.nerc.ac.uk/bio-linux/dists/unstable/bio-linux/

Now let us have a look at the original Ubuntu12.04 repositories.

By default following repositories are enabled for Ubuntu 12.04(it is called "Precise Pangolin"; Mark Shuttleworth has the tendency to give animal names...[22,23,24]): Main, Universe, Multiverse and Backports(not complete list). Let us have a short look also at the others:

Main:
provide the core of the Ubuntu12.04/Bio-Linux, like the kernel(linux3.8.x), the unity desktop, LibreOffice,...,well all what a standard system needs and the Ubuntu Core developers put their hands on.

Universe:
Yes, this is the stuff for us. The brave, the geeks, the... scientists. Most of the scientific software we use reside in universe. The packages from the Debian-Med project(you remember:...Tille), for example.

Unfortunately the Ubuntu Core developers do not put hands on this(in general).
The cause is that there is more software out there than developers, so they focus on the main repository to provide a maintained core system.

Multiverse:
Here reside the software, where is doubt if it is really free software.Sometimes it is clear that the Software is non-free. Sometimes the software only depends on non-free components.

> **By the way:**
>
> If you decide to work on the science packages in "universe" for the benefit of all of us, you can find a software QA page created by the Debian-Med team, that point also to the bugs in Ubuntu bug tracker:
>
> http://qa.debian.org/developer.php?login=debian-med-packaging@lists.alioth.debian.org&ordering=3
>
> and here is a good developer guide to get started:
>
> https://wiki.ubuntu.com/UbuntuDevelopment
>
> But beware. This is the deep look with the long journey. Nothing for a weekend, nothing for short term enthusiasm. You count here in months.

Backports:

For short: It brings newest software to you.

Ubuntu(the mother of Bio-Linux) is at most strictly oriented forward[34]. So, the big bunch of work time of the Ubuntu Core developers are going into the actual development branch. What you get in the older (final) releases, those which are still supported, is what is falling from the desk.

If a fix in the actual development fits also for a problem in one of the previous final releases the patch will also brought there. The rest of the system of final releases will be kept untouched. There is no special further development on them.

On LTS versions, like Ubuntu 12.04(It is the basis for Bio-Linux7)the support period is 5 years[20], in the style described above. No newer software will be get into Ubuntu12.04 in this way and with this not in Bio-Linux 7 throught the Ubuntu12.04 repositories.

For getting around this the backport repository was introduced.

The development of Ubuntu-Linux is a more evolutionary one. There are mostly no hard breaks. You have the opportunity to upgrade from one Ubuntu release version to another[25]. You see this more directly on the distribution upgrade capability from Bio-Linux 6 to Bio-Linux 7, based on the upgrade capabilities between Ubuntu 10.04 LTS and Ubuntu12.04 LTS[26].

And there fits backporting.

Even when you use(and you really should - more stability, better scientific validity through fixes of not obvious calculation/results errors in scientific software) the distribution upgrade, there is a period of waiting around 2 years until you can do that(until the new LTS is final).

For bridging over this gap, we have the backporting repository.

The process of backporting is driven by the community people like you and me and start with looking what is there at software for the latest development branch of Ubuntu[27].

It follows manually downloading the package from the latest development versions of Ubuntu, rebuild them with Ubuntu-dev-tools and setting up some virtual machines for testing if the rebuild packages are working on previous releases of Ubuntu. Afterwards the working packages must be uploaded to the Ubuntu servers and a report for the migration into the official

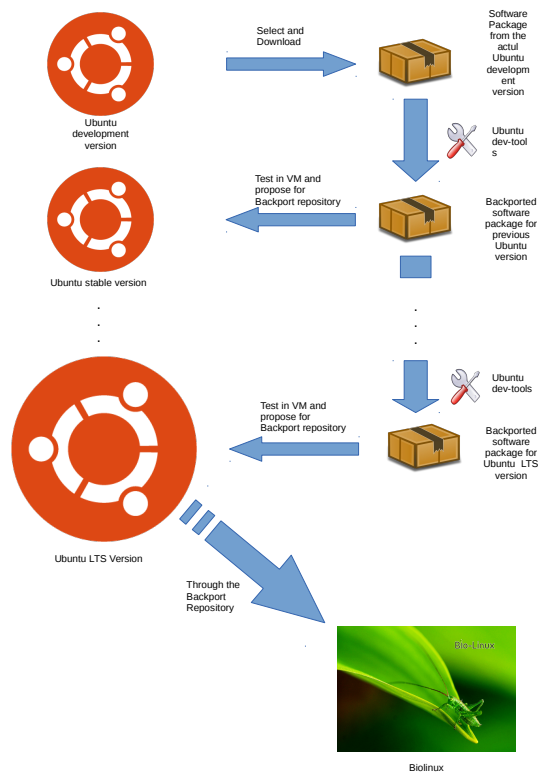

*Figure2: The scheme of the backporting workflow is shown. Icons and images are from different sources[35,36,37,38]. The graphic was designed with LibreOffice Draw4.1.2 on Ubuntu13.10*

backport repositories must be filed and maintained in cooperation with the Ubuntu(core)developers. There is a online documentation to the backporting process, which gives deeper explanation of the technical process:

https://wiki.ubuntu.com/UbuntuBackports

**What has concrete be done so far:**

There was a manually selection of bioinformatic software that was new to one of the latest normal releases (Ubuntu13.04 or Ubuntu12.10) and not available in the latest LTS version of Ubuntu, which is currently Ubuntu12.04. This gave a list of 43 programs. This selection was reduced again by substracting those programs that are available from any by default enabled software repository in Bio-Linux7. The final selection of programs was ported back to Ubuntu12.04 and over all intermediated releases between the normal releases(Ubuntu13.04 and Ubuntu12.10) and the Ubuntu LTS 12.04. This process was done with the usage of the ubuntu-dev-tools[27].

As a result there where 14 new programs, brought with the support of Felix Greyer[40] into the backporting repositories for Ubuntu 12.04 and with this into Bio-Linux 7, that are not available from any other repository that is enabled in Bio-Linux 7. See Table1 below:

| Software-Package | Debian Med ppa | Biolinux ppa | Tim Boot ppa | Ubuntu Raring | Ubuntu Quantal | Ubuntu Precise | Sonstiges: | Building: | Uploaded: | Backported: | From: |
|---|---|---|---|---|---|---|---|---|---|---|---|
| arb | n/a | 5.5-1ubuntu1 | 5.3-3ubuntu3 | 5.3-4ubuntu2 | 5.3-4ubuntu2 | n/a | Bugreport | no task | success | no task | |
| paml | n/a | n/a | n/a | 4.5-1 | 4.5-1 | n/a | | success | success | done | quantal2precise |
| agherman | n/a | n/a | n/a | 0.7.5.1-1 | 0.6.0.1-1build1 | n/a | | success | success | done | quantal2precise |
| bowtie2 | 2.0.2-1~precise1 | 2.1.0-1-0ubuntu1 | n/a | 2.0.5-1 | 2.0.0-beta6-3ubuntu1 | n/a | | no task | success | no task | |
| concavity | n/a | n/a | n/a | 0.1-2 | n/a | n/a | | success | success | done | raring2quantal |
| conservation-code | n/a | n/a | n/a | 20110309.0-1 | n/a | n/a | | success | success | done | raring2quantal |
| dssp | n/a | n/a | n/a | 2.0.4-2 | 2.0.4-2 | n/a | | failed | no task | no task | |
| fastqc | 0.10.1+dfsg-1~precise1 | n/a | n/a | 0.10.1+dfsg-1 | n/a | n/a | | failed | no task | no task | |
| fasttree | n/a | n/a | n/a | 2.1.5-1 | 2.1.4-1 | n/a | | success | success | done | quantal2precise |
| ffindex | n/a | n/a | n/a | 0.9.9-2 | 0.9.6.1-1 | n/a | | success | success | done | quantal2precise |
| Freemedforms-project | 0.8.0-precise1 | n/a | n/a | 0.7.6-1 | 0.7.6-1 | n/a | | success | success | done | quantal2precise |
| genometools | n/a | n/a | n/a | 1.4.2-4 | n/a | n/a | | success | success | done | raring2quantal |
| grinder | n/a | n/a | n/a | 0.4.5-1 | 0.4.5-1 | n/a | Bugreport | failed | no task | no task | |
| hhsuite | n/a | n/a | n/a | 2.0.15-1 | 2.0.15-1 | n/a | | failed | no task | no task | |
| hmmer2 | n/a | 2.3.2-6 | n/a | 2.3.2-6 | n/a | n/a | | no task | no task | no task | |
| king | n/a | 2.21.120420+r | 2.21.120420+r | 2.3.2-6 | n/a | n/a | | no task | no task | no task | |
| logol | n/a | n/a | n/a | 1.5.0-6 | 1.5.0-6 | n/a | | success | success | done | quantal2precise |
| ltrsift | n/a | n/a | n/a | 1.0.1-1 | n/a | n/a | | failed | | | |
| ncbi-seq | n/a | 0.0.20000620-1 | n/a | 0.0.20000620-1 | n/a | n/a | | no task | | | |
| neobio | n/a | n/a | n/a | 0.0.20030929-1 | 0.0.20030929-1 | n/a | lib | success | success | not filed | |
| norsnet | n/a | 1.0.16-1 | n/a | 1.0.16-1 | n/a | n/a | | no task | | | |
| norsp | n/a | 1.0.5-1 | n/a | 1.0.5-1 | n/a | n/a | | no task | | | |
| orthanc | n/a | n/a | n/a | 0.4.0-1 | n/a | n/a | | failed | | | |
| predictnls | n/a | 1.0.20-1 | n/a | 1.0.20-1 | n/a | n/a | | no task | | | |
| predictprotein | n/a | 1.0.86-1 | n/a | 1.0.90-1 | n/a | n/a | | no task | | | |
| profbval | n/a | 1.0.22-1 | n/a | 1.0.22-1 | n/a | n/a | | no task | | | |
| profisis | n/a | 1.0.11-1 | n/a | 1.0.11-1 | n/a | n/a | | no task | | | |
| proftmb | n/a | 1.1.10-1 | n/a | 1.1.12-1 | 1.1.10-1 | n/a | | no task | | | |
| ray | n/a | 2.2.0-0ubuntu1 | n/a | 2.1.0-1 | n/a | n/a | | no task | | | |
| reprof | n/a | n/a | n/a | 1.0.1-1 | 1.0.1-1 | n/a | | success | success | done | quantal2precise |
| saint | n/a | n/a | n/a | 2.3.3-1ubuntu2 | 2.3.3-1ubuntu1 | n/a | | success | success | done | quantal2precise |
| snappy1.0.3-java ? | 1.0.3-rc3~dfsg-3~precise1 | 1.0.4.1~dfsg-1 | n/a | 1.0.3-rc3~dfsgcn | n/a | n/a | | no task | | | |
| soapdenovo | n/a | n/a | n/a | 1.05-1 | n/a | n/a | | success | success | done | raring2quantal |
| tophat | 2.0.6-1~precise1 | 2.0.8-0ubuntu1 | 1.4.0-0ubuntu1 | 2.0.6-1 | 2.0.3-1 | n/a | | no task | | | |
| trimmomatic | n/a | n/a | n/a | 0.22-1ubuntu1 | n/a | n/a | | failed | no task | no task | |
| volview | n/a | n/a | n/a | 3.4-3ubuntu1 | 3.4-3build1 | n/a | | success | success | done | quantal2precise |
| beast-mcmc | n/a | n/a | n/a | 1.6.2-3ubuntu2 | 1.6.2-2ubuntu1 | n/a | | failed | | | |
| spread-phy | n/a | n/a | n/a | 1.0.5+dfsg-1 | n/a | n/a | | failed | | | |
| blimps | n/a | 3.9-1ubuntu1 | n/a | 3.9-1 | 3.9-1 | n/a | | no task | | | |
| cluster3 | n/a | n/a | n/a | 1.50-1 | n/a | n/a | | success | success | done | raring2quantal |
| phy-spread | n/a | n/a | n/a | 1.0.3-1ubuntu1 | 1.0.3-1ubuntu1 | n/a | | failed | | | |
| sift | n/a | 4.0.3b-5 | n/a | 4.0.3b-4 | 4.0.3b-3ubuntu1 | n/a | | no task | | | |

*Table 1: The spreadsheet shows the software selection from Ubuntu13.04 Raring done during the June 2013. Selection criteria was that the programs are not already in Ubuntu12.04 Precise and are not available through any other repository that is enabled in Bio-Linux7. If the software was already available through such a repository(ppa) the status in the column backported was set to "no task"."done" in the Backported column means that all backport steps like building and uploading are done with success, the functional test in a virtual machine of Ubuntu12.10 Quantal and Ubutnu12.04 Precise were also performed with positive result and the offical upload to the backport repositories has been sponsored by Felix Greyer[40]. The spreadsheet was designed with LibreOffice Calc4.1.2 on Ubuntu13.10.*

This all was sent to the Bio-Linux mailing list with the question to join into a for now e-mail coordinated team which do the backports for Bio-Linux[28].

**Expected backporting rounds:**

The question arises how to bring this all in a repeatable workprocess that is able to make sure that there is a regular import of the latest bioinformatic software in Bio-Linux.

A short examination of the Ubuntu release cycle shows, that the import freeze with stops the import of new packages from Debian testing or unstable...[LTS version of Ubuntu] seems to be the best time point for starting the backport process to Ubuntu12.04 and with this to Bio-Linux 7, because with this date all imports for this release version of Ubuntu should have be done. Also the bioinformatic ones[29].

So, for Bio-Linux 7 there is a just finished backporting round and there will be two further ones. One of them when Ubuntu13.10 will reach its import freeze in year 2013, and the second when Ubuntu14.04 will reach this milestone in year 2014.

After this two rounds it is to expect that Tim Booth[42] and team are going to design Bio-Linux 8 on top on Ubuntu 14.04. A migration from Bio-Linux 7 to Bio-Linux 8 will be highly recommend, because the backporting cycles planned here, will only target software that are completely not available in Bio-Linux. Software updates on this was for already in Bio-Linux residing software will not be done in respect to the system stability and to high workload for the here just founded backporter team.
You know that is all done by colleagues of you for you in their free time. They are not payed for the gift they provide to us all.

**The planned Bio-Linux 8 backporting rounds:**

For the Bio-Linux 8 cycle itself there will be backporting rounds if the following planned Ubuntu version will reach their import freez in their development cycles:

14.10, 15.04, 15.10 and 16.04

It is askable if this release cycles and with this the plans on top of them described here will be kept in this way for the future.

There is a just started discussion about the shortening of the Ubuntu release cycles[30].
In a podcast with Leann Ogasawara[43] one of the leading kernel developer for Ubuntu, she has announced to transform Ubuntu from a discrete release cycle which ends with final releases to that was is called a rolling release(see the explanation in the first part of this paper)[31].

So, Ubuntu might be become similar to that was Arch Linux already is[13], or what Greg-Kroah Hartman[44] has introduced for openSUSE with his tumbleweed repository[32, 33].
This may force a change in the plans for bio backporting described here, but the change of the Ubuntu release cycle is still under discussion. We will see what to do if and when it become finished.

**What about the future for the bioinformatic backporting?**

In moment the following steps are hand crafted, which makes them cause remarkable workloads:

- Obtaining new packages in Debian Med
- Check if this software already resides in the LTS version of Ubuntu on which Bio-Linux is constructed.
- Backporting in first steps on private server resources.
- Manually testing the functionality of the backported packages and upload to the Ubuntu project servers.
- Filing a ticket for the official backport to the public servers.

All steps listed here should be capable of becoming automated with time, except for the manually testing the functionality of a software. The last is still in a sense of full automation a non reached holy grail of informatics and a problem that will be here not tried to solved drive by.

**Conclusion:**

As could be seen here the import of the latest bioinformatic software into Bio-Linux has a inherent delay of 2 years which is caused by the design and construction process in which every Bio-Linux version is created.
With this the latest result of the scientific investigations in bioinformatics as far as algorithm and software is in view are not available to the heart of bioinformatics community. It is not hard to estimate that this circumstance has slowing down effect on the scientific progress in this area.

The other way around, every bridging over this 2 year gab must cause a speedup of the scientific progress under the setting that there a better tools available over time for doing bioinformatic jobs or tools that allow to explore in new areas that were not accessible before.
To be sure that the attention of the Bio-Linux community will get on this possibilities the Bio-Linux mailing list

bio-linux@nebclists.nerc.ac.uk

is used to inform a huge number of bioinformatic scientist at once about the latest available tools through backport, to get the latest and hopefully greatest in world wide use.